\documentclass[prl,aps,twocolumn,showpacs]{revtex4}

\usepackage{graphicx}
\usepackage{epsfig}
\usepackage{psfrag}

\newcommand{\kbar}{\mathchar'26\mkern-9muk}

\begin{document}
\title{Diffusion Resonances in Action Space\\ for an Atom Optics Kicked Rotor with Decoherence}
\author{A.J. Daley}
\author{A.S. Parkins}
\author{R. Leonhardt}
\author{S.M. Tan}
\affiliation{Department of Physics,
University of Auckland, Private Bag 92019, Auckland, New Zealand}
\date{1 August 2001}

\begin{abstract}
We numerically investigate momentum diffusion rates for the pulse kicked rotor
across the quantum to classical transition as the dynamics are made more macroscopic 
by increasing the total system action. For initial and late time rates we observe an enhanced diffusion
peak which shifts and scales with changing kick strength, and we also observe distinctive peaks around quantum resonances.
Our investigations take place in the context of a system of ultracold atoms which is coupled to its environment via spontaneous emission
decoherence, and the effects should be realisable in ongoing experiments.
\end{abstract}

\maketitle

The transition from quantum to classical behaviour in nonlinear dynamical systems
has been a topic of much interest in recent years, motivated by the large differences that exist between the behaviour of such systems
in the two regimes. The $\delta$-Kicked Rotor(DKR) is a prime example - in particular, coherence effects
in the quantum DKR completely suppress classical diffusion \cite{scott1,scott2}. The 
quantum DKR is also very interesting because it has been beautifully demonstrated in experiments which probe the 
momentum distribution of a cloud of laser-cooled atoms interacting with a 
pulsed standing wave of near resonant light \cite{scott3}. 
These experiments necessarily involve a generalisation of the DKR to kicks of finite length, and it is this system,
the Kicked Rotor(KR) which we investigate in this paper.

There have been numerous studies, both theoretical (see for example \cite{scott4,scott5,scott6,scott7,scott8}) 
and experimental \cite{scott9,scott10,scott11,scott12,scott13}, of the role of decoherence in the 
quantum to classical transition for the Kicked Rotor. As with any real quantum system, the atom optics KR
couples to its environment, resulting in a loss of phase coherence. In the case we consider here,
this coupling is between the atoms and the vacuum electromagnetic field, and results in atomic spontaneous emissions and
concomitant random momentum recoils.

Most work in the past has focussed on changing the level of this decoherence and observing the effects on momentum 
diffusion rates and distributions, looking in particular at how increased levels of decoherence
``drive'' the quantum system towards classical behaviour. In this paper we focus instead on what happens 
when we fix the level of decoherence in our real quantum system, and then make the dynamics more 
macroscopic by varying the total action in the system; that is, by varying the effective Planck's constant. 
In so doing we find dramatic structures in the momentum diffusion rates, similar to those found recently for the DKR 
with a continuous position measurement \cite{KJS+00}, only here we consider a specific experimental configuration.

The system we model is a cloud of ultracold Caesium atoms (of initial temperature $\approx 10\mu K$) which interact with
a standing wave of laser light of frequency $\omega_l$, detuned far from the frequency $w_0$ of the 
$6S_{1/2} \rightarrow 6P_{3/2}$ atomic transition. The laser is pulsed with period $T$ and pulse profile $f(t)$.
If the detuning is large, the internal atomic dynamics can be eliminated, and the resulting single particle 
Hamiltonian (for just the external degrees of freedom) is \cite{scott3}
\begin{equation}
\hat{H}=\frac{\hat{p}^2}{2m}-\frac{\hbar \Omega_{eff}}{8} \cos(2k_l \hat{x})\sum^{N}_{n=0} f(t-nT) ,
\end{equation}
where $\hat{x}$ and $\hat{p}$ are the atomic position and momentum operators, respectively, and $k_l$ is the wave
number of the laser light. The effective potential strength, 
$\Omega_{eff}=\Omega^2(s_{45}/\delta_{45}+s_{44}/\delta_{44}+s_{43}/\delta_{43})$, accounts 
for the different dipole transitions between hyperfine levels in the Caesium atoms 
(F=4 $\rightarrow$ F'=5,4,3), with  $\delta_{ij}$ the corresponding detunings, and $\Omega/2$ the (single-beam)
resonant Rabi frequency. If we assume equal populations in all Zeeman sublevels, then $s_{45}=\frac{11}{27}$, 
$s_{44}=\frac{7}{36}$, and $s_{43}=\frac{7}{108}$. We can rewrite this Hamiltonian in appropriate dimensionless 
units as
\begin{equation}
 \hat{H}'=\frac{\hat{\rho}^2}{2} - k \cos{\hat{\phi}} \sum^{\infty}_{n=0}  f(t'-n) , 
\end{equation}
which is the Hamiltonian for the standard kicked rotor system.
Here, $\hat{\phi}=2k_l\hat{x}$, $\hat{\rho}=2k_lT\hat{p}/m$, $t'=t/T$, and $\hat{H}'=(4k_l^2T^2/m)\hat{H}$.
The classical stochasticity parameter is given by $\kappa=\Omega_{eff}\omega_RT\tau_p$, where $\tau_p$ is the pulse
length and $\omega_R=\hbar k_l^2/2m$. In our work $f(t')$ is generally a square pulse, i.e., $f(t')=1$ for $ 0<t'<\alpha$,
where  $\alpha=\tau_p/T$. Note that $k=\kappa/\alpha$.
In these units, we have $[\hat{\phi},\hat{\rho}]=i\kbar$, with $\kbar=8\omega_RT$. Thus the quantum 
nature of the system is reflected by an effective Planck's constant, $\kbar$, which scales as we change the total action 
in the system by altering the pulse period $T$.

Decoherence occurs in the form of spontaneous emission events, which occur when the atoms absorb
light from the standing wave \cite{scott9}. It is assumed that momentum distributions in orthogonal directions
remain independent, and thus the system remains effectively one dimensional.
We characterise the level of this decoherence by the probability of spontaneous emission
per kick, $\eta$. Given the large detuning, i.e., $\Omega_{eff}/\delta \ll 1$, this process may
be modelled by the master equation for the density operator $\hat{w}$ of the system \cite{scott8} 
\begin{eqnarray}
\label{master1}
\dot{\hat{w}}&=&-i [\hat{H},\hat{w}]-\frac{\eta}{\alpha}\sum^{N}_{n=0} f(t-n)[\cos^2(\hat{\phi}/2),\hat{w}]_+\nonumber \\
& &+ 2\frac{\eta}{\alpha}\sum^N_{n=1}f(t-n)\int^1_{-1}du N(u)e^{i u \hat{\phi}/2} \nonumber \\
& &\times \cos(\hat{\phi}/2)\hat{w}\cos(\hat{\phi}/2)e^{-i u \hat{\phi}/2} ,
\end{eqnarray}
where $N(u)$ is the distribution of recoil momenta projected onto the standing wave axis, and $[.,.]_+$ denotes an anti-commutator.
We have also modelled spontaneous emission events in which the atoms absorb light from 
counterpropagating beams of oppositely circularly polarised light, 
which interact with the atoms continuously \cite{scott10}. This leads to emission events which are independent of
position, and which may occur at times other than during a kick. 
The results obtained for the two types of spontaneous emission noise are very similar, and so we only present 
the results for the first type here.

\begin{figure}[hbt]
\psfrag{kbar}[0][0][1][0]{$\kbar$}
\includegraphics[width=6.6cm]{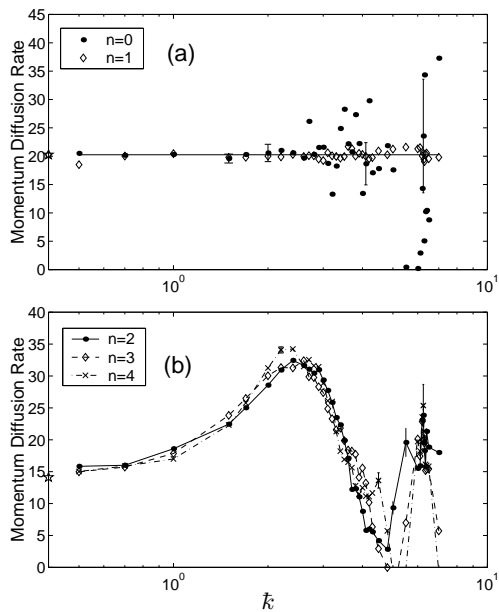}
\caption{Momentum diffusion rates $D(n)$ in the quantum kicked rotor as a function of the effective Planck's constant, $\kbar$ 
for (a) the first two kicks ($n=0$ and $n=1$) and (b) the third, fourth and fifth kicks ($n=2$, $n=3$ and $n=4$), 
with $\eta = 10\%$, $\kappa=9$, and $\alpha=0.005$. Classical values are marked on the vertical axis.}
\label{idr1}
\end{figure}

We simulate Eq.(\ref{master1}) using the method of quantum trajectories, as described in Ref. \cite{scott8}. We
choose initial momentum eigenstates from a Gaussian distribution of width $\sigma_{\rho}/\kbar=\sigma_p/2\hbar k_l=4$ 
(which reflects the initial temperature distribution) and we take an incoherent average over the final momentum
distributions. We choose $\alpha$ to be small so that the effects of KAM boundaries are not important \cite{scott9}.
Typically we use 1000 trajectories, and we calculate statistical errors (shown for some points in the figures) based on 
dividing these trajectories equally into 10 groups and computing errors in the means.
We are primarily interested in the momentum diffusion rate, which is defined as the change in the 
kinetic energy from one kick to the next, $2D(n)=<\hat{\rho}_{n+1}^2>-<\hat{\rho}_n^2>$,
where we denote $\hat{\rho}_0=\hat{\rho}(t'=0)$, $\hat{\rho}_1=\hat{\rho}(t'=1)$ etc.

Our simulation results for the momentum diffusion rates across the first five kicks are shown in Fig. \ref{idr1}. Aside 
from some noise in the simulations for larger values of $\kbar$, we see that for the first two kicks
the diffusion rates are essentially constant with respect to $\kbar$. This is the quantum version of the 
quasilinear behaviour known to exist classically in the kicked rotor system. In fact one can show (for the DKR) 
that for a uniform initial distribution of positions, $2D(0)=\kappa^2/2$ (shown as a solid line in Fig. \ref{idr1}(a)). 
Similarly, if we also assume an initial Gaussian momentum distribution of standard deviation $\sigma_\rho$, 
it can be shown that
\begin{eqnarray}
2D(1) &=& \frac{1}{2}\kappa^2 (1-J_2(K_{2q})e^{-2\sigma_\rho^2})-2\kappa J_1(K_q)\sigma_\rho^2e^{-\frac{\sigma_\rho^2}{2}}\nonumber\\
& &+\kappa^2(J_0(K_q)-J_2(K_q))\cos(\kbar/2)e^{-\frac{\sigma_\rho^2}{2}},
\end{eqnarray}
where $J_n$ is an ordinary Bessel function of order $n$, $K_q=2\kappa \sin(\kbar/2)/ \kbar$ and $K_{2q}=2 \kappa \sin(\kbar)/ \kbar$.
For sufficiently large $\sigma_\rho$, this reduces to the same result as for the first kick, which can be seen for our system 
in Fig. \ref{idr1}(a).

\begin{figure}[htb]
\psfrag{kbar}[0][0][1][0]{$\kbar$}
\includegraphics[width=6.5cm]{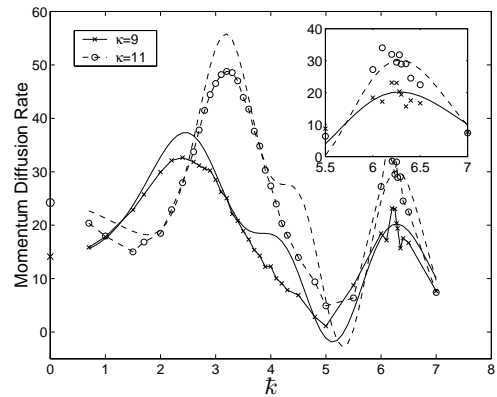}
\caption{Initial quantum diffusion rates in the kicked rotor for varying values of 
$\kappa$ and $\kbar$, with $\eta=10\%$ and $\alpha=0.005$. The points show simulation
results (an average over $D(2-5)$), while the lines show Shepelyansky's formula, Eq.(\protect{\ref{shepform}}).
Classical values are marked as points for $\kbar=0$. Note the use of a linear scale for $\kbar$.}
\label{idrshep}
\end{figure}

After the second kick, the system settles down into its initial quantum diffusion period, where for a small time
the system exhibits classical-like diffusion, with a relatively constant momentum diffusion rate, $D_q$. As can be seen
from Fig. \ref{idr1}(b), the $\kbar$ dependence of this rate is quite remarkable. We observe an enhanced diffusion
peak (or resonance) around $\kbar=3$ which shifts to the right and increases in magnitude as we increase $\kappa$ (See Fig. \ref{idrshep}).
There is also a peak in the diffusion rates near the quantum resonance at $\kbar=2\pi$ \cite{scott16}. (This structure also
occurs at larger multiples of $2\pi$). 
The quantum diffusion rates that we observe in this regime agree well with those predicted by
Shepelyansky under the conditions $\kbar \geq 1$ and $\kappa \gg \kbar$ \cite{scott14}, i.e.,
\begin{equation}
\label{shepform}
D_q=\frac{\kappa^2}{2}\left(\frac{1}{2}-J_2(K_q)-J_1^2(K_q)+J_2^2(K_q)+J_3^2(K_q)\right).
\end{equation}
This can be seen in Fig. \ref{idrshep}, where we plot the average of the curves for $D(2-5)$.
There is surprisingly good agreement for the quantum resonance peak and for the position of 
the enhanced diffusion peak, especially considering that the condition $\kappa \gg \kbar$ does not hold for our large
$\kbar$ values.
 The discrepancy in the height of the peak is created mainly by our choice to average
over the diffusion rates from 4 different kicks. It is a qualitative decision as to
when the system has really settled into the initial quantum diffusion regime, but averaging over $D(2-5)$ as we do 
gives us an objective estimate of the corresponding diffusion rate. However, in some cases (particularly 
near the maximum of the enhanced diffusion peak), the diffusion rate will already have begun to decrease 
towards the late time diffusion regime before the sixth kick.

The initial quantum diffusion period lasts for a small number of kicks, after which the diffusion rate begins to
decrease. In the absence of noise, the system settles into a localised state \cite{scott3},
and $D(n) \rightarrow 0$ as $n\rightarrow \infty$. However, the onset of dynamical localisation is a coherence effect, 
and in the presence of decoherence the system settles into a late time diffusion regime where
$D(n) \rightarrow D_{\infty} \neq 0$ as $n \rightarrow \infty$.

\begin{figure}[htb]
\psfrag{kbar}[0][0][1][0]{$\kbar$}
\includegraphics[width=6.5cm]{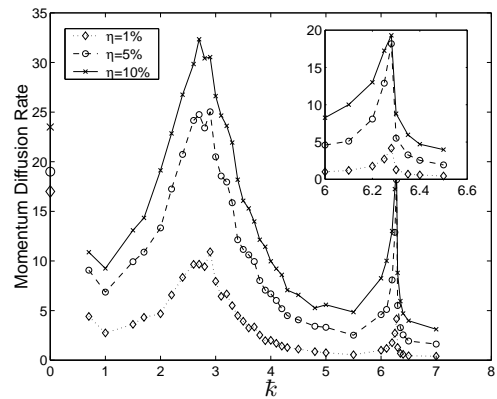}
\caption{Late time momentum diffusion rates in the quantum kicked rotor for varying values of $\eta$ and $\kbar$, with 
$\kappa=10$ and $\alpha=0.005$. The insert shows the peak near the quantum resonance at $\kbar=2\pi$, and classical values
are marked for $\kbar=0$. Note the use of a linear scale for $\kbar$.}
\label{varyeta}
\end{figure}

\begin{figure}[hbt]
\psfrag{kbar}[0][0][1][0]{$\kbar$}
\includegraphics[width=6.5cm]{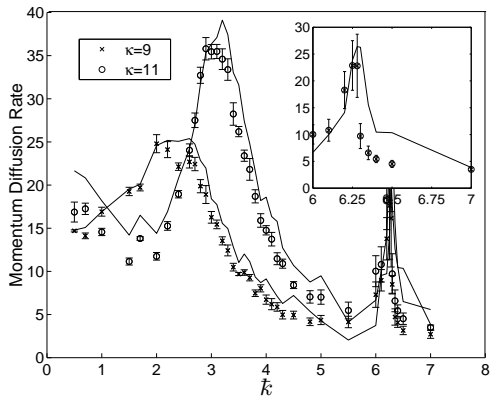}
\caption{Graph showing simulated late time momentum diffusion rates (points) and results from Eq.(\protect{\ref{sumldr}}) 
(solid lines), with $\eta=10\%$ and $\alpha=0.005$. 
The insert shows the peak near the quantum resonance at $\kbar=2\pi$ for $\kappa=11$.}
\label{latecomp}
\end{figure}

Our simulation results for these late time diffusion rates are shown for varying levels of decoherence, $\eta$, in 
Fig. \ref{varyeta}, and for varying $\kappa$ in Fig. \ref{latecomp}. Again we observe an enhanced diffusion peak (or resonance) which
shifts and scales with increasing $\kappa$, as well as a much more narrow peak near the quantum resonance at 
$\kbar=2\pi$. The most notable feature here is that for appropriate values of $\kappa$ and $\eta$, the momentum diffusion
rates near the top of the enhanced diffusion peak are actually larger than the corresponding classical values (marked on the graph),
even if
we account for added momentum diffusion due to spontaneous emission in the classical system.

The late time diffusion rates may be approximated by the formula \cite{scott6,scott9}
\begin{equation}
\label{sumldr}
D_{\infty}=\sum^{\infty}_{n=0}\eta(1-\eta)^nD_0(n),
\end{equation}
where $D_0(n)$ is the diffusion rate at the $n$th kick for a KR \textit{without} decoherence.
Essentially, we assume here that dynamical correlations over particular time intervals are suppressed by a factor 
expressing the probability that a spontaneous emission occurs within that time interval. The correlations taken over a set number of kicks
give rise to the diffusion rates seen in the KR without decoherence after that number of kicks. Hence, we take a weighted average 
over the diffusion rates as the KR settles down, where the weighting for $D_0(n)$ gives the probability that the first spontaneous emission 
event occurs during kick number $n+1$, i.e., $\eta(1-\eta)^n$. The early time diffusion rates are thus ``locked in'' by the loss
of phase coherence.


In Fig. \ref{latecomp} the simulation results for $D_{\infty}$ are plotted as points, and calculations of the 
right hand side of Eq.(\ref{sumldr}) based on calculations of $D_0(n)$ from simulations of the KR 
without decoherence are shown as solid lines. There is very good agreement between the two sets of values, especially
for the enhanced diffusion peak. Note that the values given by the solid lines contain statistical
errors from the simulations of $D_0(n)$ which are comparable in magnitude to those displayed in the figure.
The level of agreement indicates 
that the model associated with Eq.(\ref{sumldr}) works very well for the late time diffusion rates.

Late time diffusion rates greater than the corresponding classical rates occur because quantum correlations 
cause the initial quantum diffusion rates to be higher than the corresponding classical rates in the appropriate 
cases, and these higher rates become locked in by the loss of phase coherence. These decoherence effects are thus 
much more important than the more direct increase in momentum diffusion due to the recoil in a spontaneous 
emission process which may be accounted for classically as well as quantum mechanically.

\begin{figure}[h!bt]
\psfrag{kbar}[0][0][1][0]{$\kbar$}
\includegraphics[width=6.5cm]{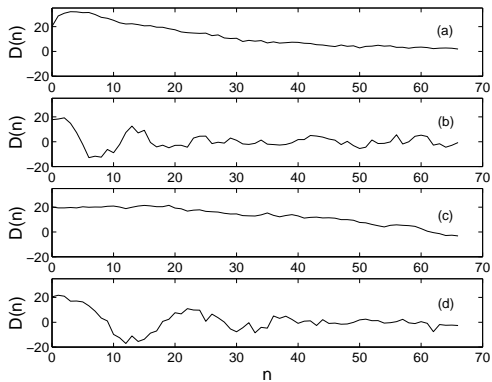}
\caption{Graph showing diffusion rates as a function of kick number for (a) $\kbar=2$, (b) $\kbar=6$, 
(c) $\kbar=6.28$ and (d) $\kbar=6.4$, with
$\kappa=9$ and $\alpha=0.005$. Notice the inital quasilinear behaviour followed by exponential settling
for lower $\kbar$ values which contrasts with the oscillatory behaviour for $\kbar$ values near the quantum resonance
peak.}
\label{dsettle}
\end{figure}

It is possible to find an analytical expression for $D_{\infty}$ which agrees well with the simulation results over  
a large range of $\kbar$ values by making assumptions about the form of $D_0(n)$. For example, we can assume that 
$D_0(0)=D_0(1)=\kappa^2/4$, and that for $n\geq2$ the diffusion rate starts at the initial quantum diffusion rate, $D_q$, and
decays to zero exponentially with a time constant which depends 
on the quantum break time \cite{scott6}. The main problem is in determining a form that works well near the quantum
resonance, where the assumption of exponential relaxation in $D_0(n)$ breaks down and oscillations occur in the diffusion rate, as
shown in Fig. \ref{dsettle}.

The behaviour which we observe in the late time rates across the quantum to classical transition for the atom 
optics kicked rotor is similar to that observed by Bhattacharya \textit{et al.}\cite{KJS+00} for the DKR with a continuous
position measurement, and leads to similar questions about the nature of the quantum to classical transition.
The observation of such results from a real decoherence process is very interesting, and our simulations suggest 
that these results should be readily observable in laboratory experiments. In fact, hints of unusual behaviour in the 
momentum diffusion rates as a function of $\kbar$ have already been observed in experiments with cold atoms
\cite{scott17,scott18,scott19}.

We thank Kurt Jacobs for interesting and stimulating discussions and Andrew Doherty for providing the computer source 
code of Ref. \cite{scott8}, which formed the basis for our simulations. 
This work was supported by a grant (UOA016) from the Marsden Fund of the Royal Society of New Zealand.

Note added: After the completion of this work we learned of a cold atom KR experiment
[M. B. d'Arcy \textit{et al.}, Phys. Rev. Lett. \textbf{87}, 074102 (2001)]
demonstrating enhanced diffusion resonances of the sort considered in this paper.

\bibliographystyle{revtex}

\end{document}